\documentclass[aps,prc,reprint,onecolumn,superscriptaddress,nofootinbib]{revtex4-2}

\usepackage{graphicx}
\usepackage{amsmath}
\usepackage{booktabs}
\usepackage{xcolor}
\usepackage{hyperref}
\hypersetup{colorlinks=true,linkcolor=blue,citecolor=blue,urlcolor=blue}
\graphicspath{{figs/}}
\usepackage[section]{placeins}

\begin{document}

\title{FUSION: a skill-based research agent for publicly obtainable nuclear-physics codes}

\author{Jin Lei}
\email[Contact author: ]{jinl@tongji.edu.cn}
\affiliation{School of Physics Science and Engineering, Tongji University, Shanghai 200092, China.}
\affiliation{Southern Center for Nuclear-Science Theory (SCNT), Institute of Modern Physics, Chinese Academy of Sciences, Huizhou 516000, Guangdong Province, China.}

\date{\today}

\begin{abstract}
Running an unfamiliar nuclear-physics code is rarely difficult because of the physics alone. One must find and build the program, learn its input conventions, and decide whether a plausible output is actually correct. A general-purpose coding agent helps with the first two tasks but may make the last one harder: it can write an input file that runs with the wrong physical convention. FUSION addresses this problem with code-specific skills, delivered as a plugin for the coding agent a user already runs. A skill obtains the code from its public source, starts from a verified input, runs and parses the calculation, records known failure modes, and must reproduce a stated benchmark to a stated tolerance before reporting a result. The current release covers twenty codes, spanning optical models and reactions, nuclear structure, fission and statistical models, astrophysics and R-matrix analysis, and heavy-ion transport, together with the BAND framework for Bayesian calibration, emulation and model mixing. It also includes an offline, searchable collection of 61\,167 pages derived from the nucl-th literature. User notes and credentials remain outside the public repository. FUSION is available under the MIT license at \url{https://github.com/jinleiphys/FUSION}; documentation is at \url{https://vibeinscience.com}. Here I describe the design, the checks behind the current release, and one complete calculation from input to comparison with measured data.
\end{abstract}

\maketitle

\section{A calculation that runs can still be wrong}

The first calculation with an established nuclear-physics code often takes longer than the calculation itself. The source may live on a personal web page, the program may require an old compiler or library, and its input conventions may be documented only in a long manual. Much of this knowledge is learned once within a group and then lost or passed on informally.

\begin{figure}[!htb]\centering
\includegraphics[width=0.92\textwidth]{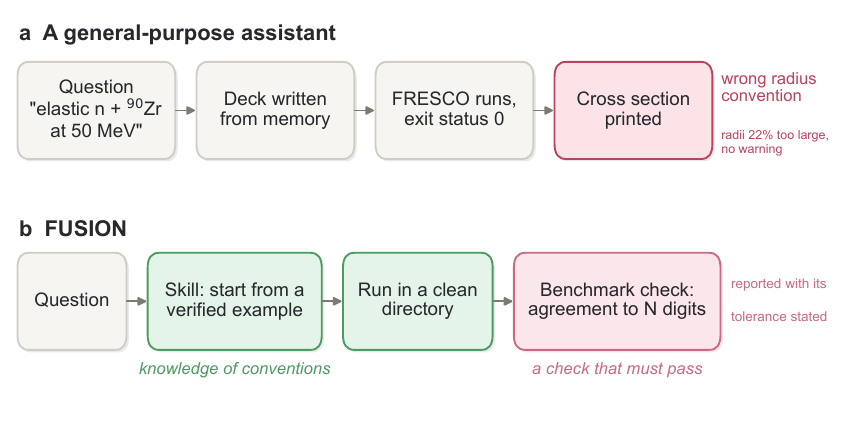}
\caption{\label{fig:problem}The failure mode and the fix. (a) A general-purpose assistant writes the input from memory; the code runs and prints a number, and a wrong convention is invisible. (b) A FUSION skill starts from a verified example deck and ends with a benchmark check whose agreement is stated in significant figures.}
\end{figure}

Language-model agents can compile programs, search manuals, and draft input files. They do not, by themselves, know whether the calculation represents the intended physics. This produces a particularly dangerous failure (Fig.~\ref{fig:problem}a): the input is syntactically valid, the program exits normally, and the output looks reasonable. In the example of Sec.~\ref{sec:run}, a general-purpose assistant writes a FRESCO~\cite{Thompson1988} input file (a deck, in the usage of that community) with the wrong radius convention. Every radius is then 22\% too large, but the program still prints a smooth angular distribution and no warning.

FUSION adds two controls (Fig.~\ref{fig:problem}b). The agent modifies a verified input instead of inventing one, and it must pass a numerical check before reporting the result. These controls do not replace physical judgment. They make silent mistakes easier to detect.

\section{What FUSION is}

\begin{figure}[!htb]\centering
\includegraphics[width=0.92\textwidth]{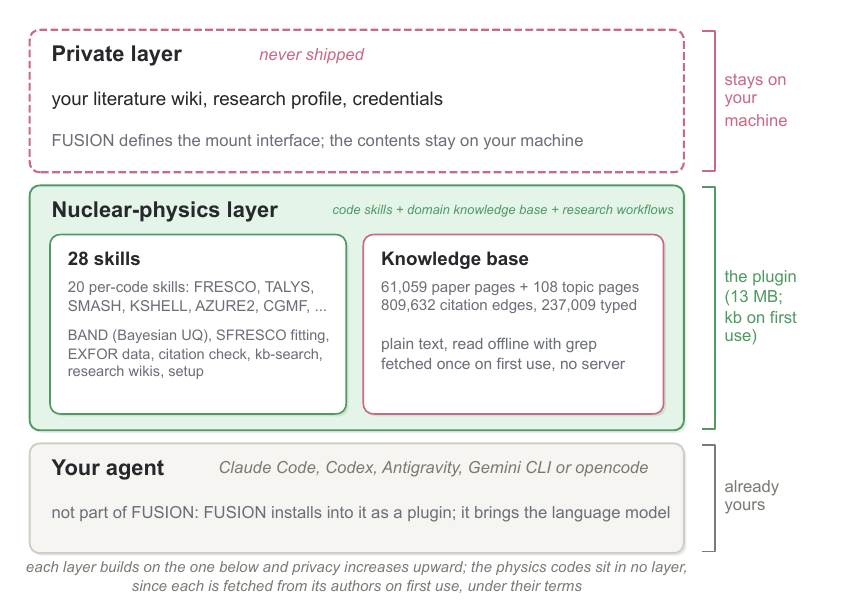}
\caption{\label{fig:arch}FUSION on the user's own agent. The nuclear-physics layer is the plugin, with the literature collection fetched on first use; the private layer is a mount point the platform defines but never fills. The physics codes belong to no layer and are obtained from their authors.}
\end{figure}

FUSION is not an agent of its own. It sits on the coding agent the user already runs, which brings the language model, and adds two layers to it (Fig.~\ref{fig:arch}), ordered so that the public software stays separate from the user's private material. It installs as a plugin; the same repository is packaged for Claude Code, Codex, Antigravity, the Gemini command line and opencode. The nuclear-physics layer contains 28 skills and the literature collection. Twenty skills operate individual physics codes; of the remaining eight, one operates the BAND framework for Bayesian uncertainty quantification~\cite{Phillips2021}, one is a fitting front end, one retrieves measured data from the EXFOR library~\cite{Otuka2014}, one queries the knowledge base, one checks citations against the publisher record, two maintain the user's own research notes, and one sets FUSION up. The plugin is about 13~MB; the literature collection, about 250~MB to download, is fetched the first time it is needed and checked against a manifest before use.

The user's literature notes, research profile, and credentials form a separate private layer. FUSION defines where these files may be attached but does not create or distribute them. The public repository therefore works without access to a user's notes, and cloning the repository cannot expose them.

The skills are the project, and none of them depends on which agent carries it. An earlier release bundled a rebranded copy of the open-source \texttt{opencode} agent for users with no agent installed; it is no longer maintained, because each agent listed above loads the skills directly. The physics codes themselves are not redistributed. FUSION reuses an installed copy or obtains the source from the authors' public site under the code's own license.

Installation uses the agent's own plugin command; in Claude Code, for example,
\begin{quote}\small\ttfamily
claude plugin marketplace add jinleiphys/FUSION\\
claude plugin install fusion@fusion
\end{quote}
The repository's README gives the equivalent for each of the other agents, written so that an agent reading it can carry out the installation itself. FUSION contains no model and asks for no key. The first request triggers an optional setup dialogue (areas of work, private space) that never repeats.

\section{What a skill is}

\begin{figure}[!htb]\centering
\includegraphics[width=0.72\textwidth]{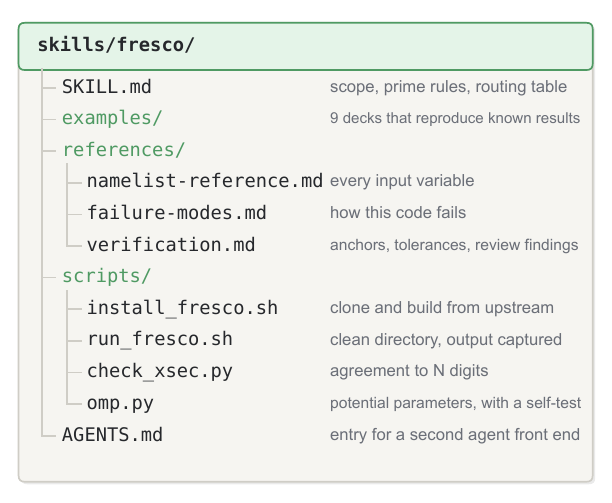}
\caption{\label{fig:anatomy}Anatomy of a skill, using FRESCO as the example. The layout is typical, not fixed: a scope document, verified examples, reference notes, and scripts that install, run and check, with the potential-parameter script carrying its own self-test.}
\end{figure}

A skill is a directory of instructions, examples, and small scripts. It contains no language model. When a request names a supported code or calculation, the agent reads the corresponding skill before acting. Figure~\ref{fig:anatomy} shows the FRESCO example. Six things are in each one.

\emph{Installation from upstream.} The skill fetches the code from wherever its authors publish it, at a pinned version, and builds it. Where a platform needs a patch, the patch is in the skill together with evidence that it changes portability and not physics.

\emph{Input from verified examples.} The skill never asks the agent to write a deck from memory. It carries working inputs that reproduce known results and instructs the agent to start from the closest one. Of the six, this one matters most: the failure it defends against is syntactically valid nonsense.

\emph{Execution.} A wrapper runs the code in a clean directory, captures the output, and enforces conventions the code needs but does not check for itself.

\emph{Output parsing.} Which file holds which observable, in what units, with what sign convention.

\emph{Failure modes.} A written record of how this particular code fails, which is where most of the practical knowledge lives. Three examples from the current set: TALYS exits with status zero even after a fatal error, so its exit code must never be trusted, and its makefile source glob is locale-dependent and silently drops thirteen files unless the locale is fixed. GiBUU reads its random seed from the first \texttt{\&initRandom} namelist, not from the first line mentioning a seed, so a wrapper that greps for the latter can report a fixed seed while the run is seeded from the clock. A recompilation of pikoe with a different compiler fails on stale module files with a message that names neither the cause nor the fix.

\emph{A benchmark with a stated tolerance.} The skill reproduces a known result and reports the agreement in significant figures. This test distinguishes an executable procedure from documentation alone.

\section{What is covered}

\begin{figure}[!htb]\centering
\includegraphics[width=0.92\textwidth]{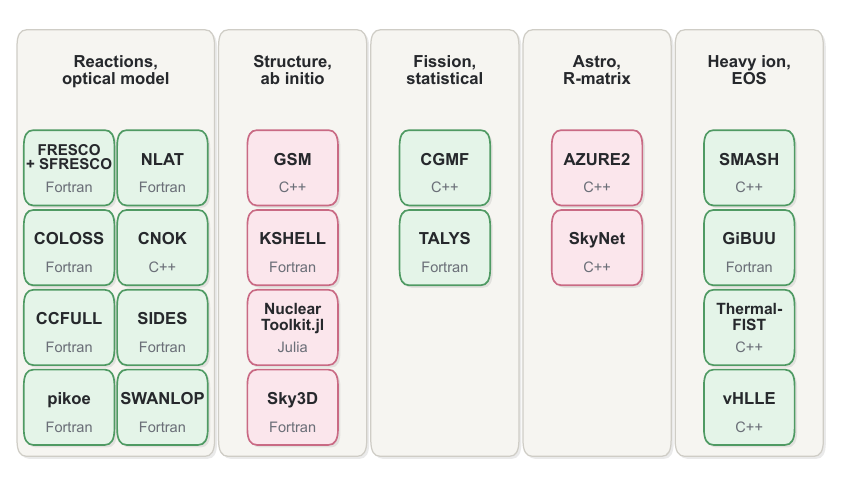}
\caption{\label{fig:coverage}The twenty codes with a FUSION skill, grouped by area, with the language each is written in. The selection is bounded by an openness rule, not by a ranking: a code qualifies only if it is publicly obtainable, builds from source on the target platform, and has a published paper.}
\end{figure}

Figure~\ref{fig:coverage} shows the twenty codes by area: FRESCO~\cite{Thompson1988} with its fitting front end SFRESCO, COLOSS~\cite{Liu2025}, CCFULL~\cite{Hagino1999}, pikoe~\cite{Ogata2024}, NLAT~\cite{Titus2016}, CNOK~\cite{Sun2023}, SIDES~\cite{Blanchon2020} and SWANLOP~\cite{Arellano2021} for reactions and optical models; GSM~\cite{Michel2021}, KSHELL~\cite{Shimizu2019}, NuclearToolkit.jl~\cite{Yoshida2022} and Sky3D~\cite{Maruhn2014,Schuetrumpf2018} for structure; CGMF~\cite{Talou2021} and TALYS~\cite{Koning2023} for fission and statistical models; AZURE2~\cite{Azuma2010} and SkyNet~\cite{Lippuner2017} for astrophysics; SMASH~\cite{Weil2016}, GiBUU~\cite{Buss2012}, Thermal-FIST~\cite{Vovchenko2019} and vHLLE~\cite{Karpenko2014} for heavy-ion collisions and the equation of state.

Coverage follows an access rule, not a ranking. A code is included only if its source can be obtained without registration, can be built on a supported platform, and is described in a published paper. The repository records why other candidates were omitted. For example, one required a language with no native Apple-Silicon toolchain, and no public source could be located for another. Omission says nothing about the scientific value of a code.

\section{How a skill is verified}

A released skill carries the evidence needed to inspect it. The record identifies the upstream revision, pinned where the upstream allows it, the tested platforms, the benchmark, and the tolerance. The checks themselves are mechanical: the code is built on both macOS on ARM and Linux on x86-64 for most skills, with the exceptions stated in their records; the benchmark is reproduced and the agreement stated in significant figures; and each guard is disabled in turn to confirm that exactly its own test fails. A second language model, different from the one that drafted the skill, then goes through it, and it may run the code as well as read it. What it finds is written into the skill and ships with it.

\begin{figure}[!htb]\centering
\includegraphics[width=0.92\textwidth]{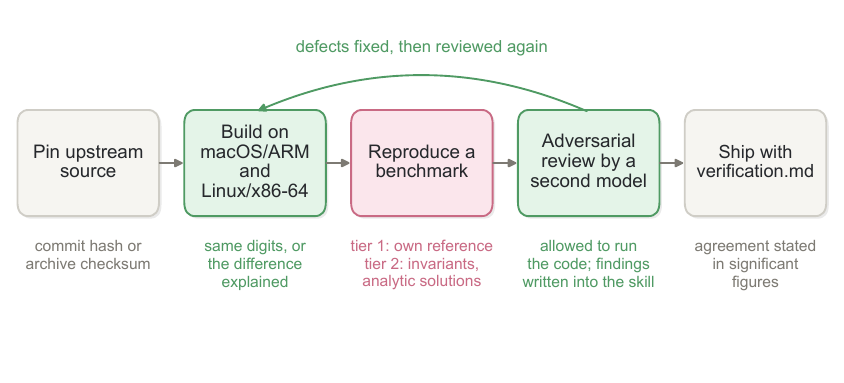}
\caption{\label{fig:pipeline}How a skill is verified before it ships. The adversarial review loops back to the build until a round returns no new defect of the kind the previous round repaired.}
\end{figure}

\begin{figure}[!htb]\centering
\includegraphics[width=0.55\textwidth]{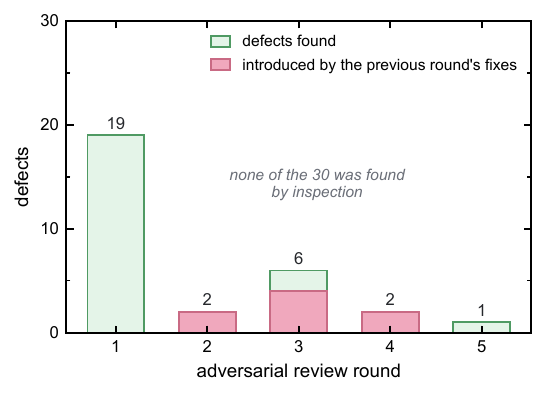}
\caption{\label{fig:rounds}Adversarial review of one skill (SMASH), round by round. Pink marks defects introduced by the fixes of the previous round. The review stops when a round returns without a new defect of that kind.}
\end{figure}

The review changes the released material. For the SMASH skill (Fig.~\ref{fig:rounds}), the first round found 19 defects, and three of the next four found further defects introduced while repairing earlier ones. What found them, in order of yield, was running the test scripts on a second machine, switching guards off one at a time, and letting the reviewer run the real binary. None of the 30 was found by inspection, including the author's own inspection of the scripts immediately after writing them, which is an argument for the mechanical checks rather than for the reviewer. Review stopped only when a complete round found no new defect of the same class.

\begin{figure}[!htb]\centering
\includegraphics[width=0.95\textwidth]{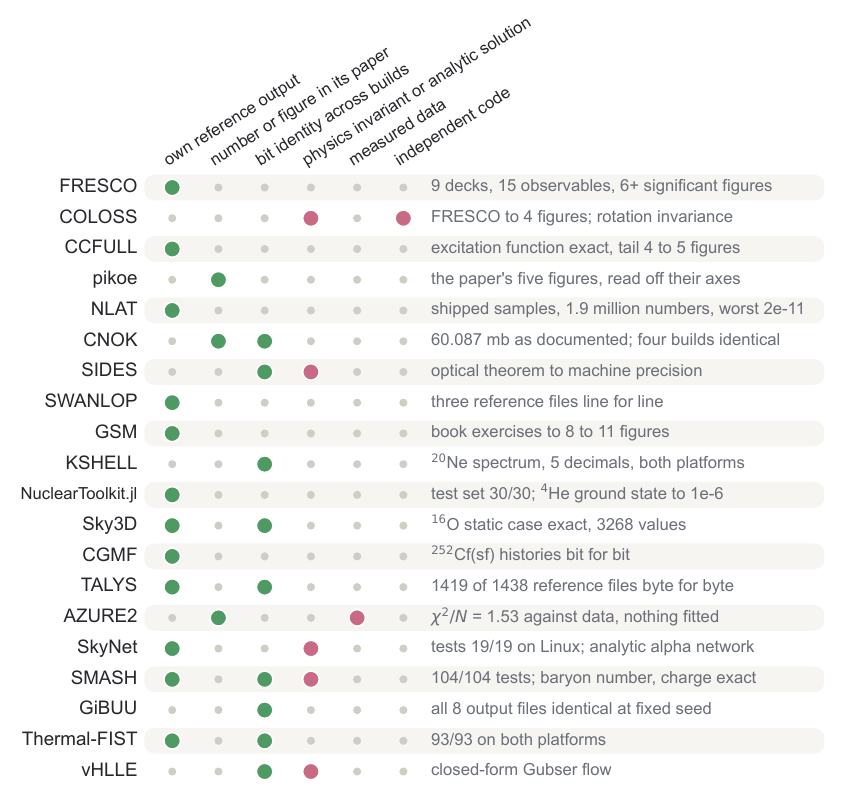}
\caption{\label{fig:pinning}What each skill's benchmark rests on. Green: reproduction of a stored result (the code's own reference, a published number, or bit identity across builds). Pink: a check that does not depend on any stored output (a physics invariant, an analytic solution, measured data, or an independent code). The right column gives the recorded agreement.}
\end{figure}

The benchmark evidence differs by code (Fig.~\ref{fig:pinning}). The repository labels each skill by its own rule: tier~1 when the code itself documents a reference value and the skill reproduces it to a stated precision (fourteen skills), tier~2 when the benchmark had to be built from something else (six). Twelve skills reproduce output distributed with the code. On macOS/ARM, TALYS matches 1419 of its 1438 reference files byte for byte; of the other 19, one differs only in the run time it records and 18 agree to about six significant figures. The same case on Linux reproduces the reference to about 5.4 significant figures. Both numbers name their platform, because a Linux user reading the first one alone would read a correct run as a broken one. CGMF reproduces the distributed $^{252}$Cf spontaneous-fission histories bit for bit. FRESCO matches fifteen observables from nine official examples covering elastic and inelastic scattering, transfer, coupled reaction channels, breakup, and capture. The other eight rest on something else, two of them still tier~1 because the value they reproduce is documented by the code's authors: published values (pikoe and CNOK), an analytic solution (vHLLE), the optical theorem (SIDES), measured data without fitted parameters (AZURE2), an independent solver (COLOSS), bit-level agreement across platforms and compilers (GiBUU), and a spectrum reproduced to every printed decimal across two platforms and two compilers (KSHELL).

Two limits need stating plainly.

\emph{A benchmark certifies that a build reproduces a known result. It does not certify that your calculation is right.} Reproducing a code's reference output proves the installation is sound and the skill drives it correctly. It says nothing about whether the model suits your problem, whether the parameters are sensible, or whether the answer means what you think. That judgment stays with the physicist.

\emph{An author's own reference output certifies build integrity, never physics,} because it was produced by the same source. A genuine physics bug is present in the reference too, and matching it proves nothing. For that reason the skills without a shipped reference rest on invariants, analytic solutions, measured data or cross-platform identity, and cross-build agreement is measured even where a reference exists.

\section{The literature layer}

\begin{figure}[!htb]\centering
\includegraphics[width=0.92\textwidth]{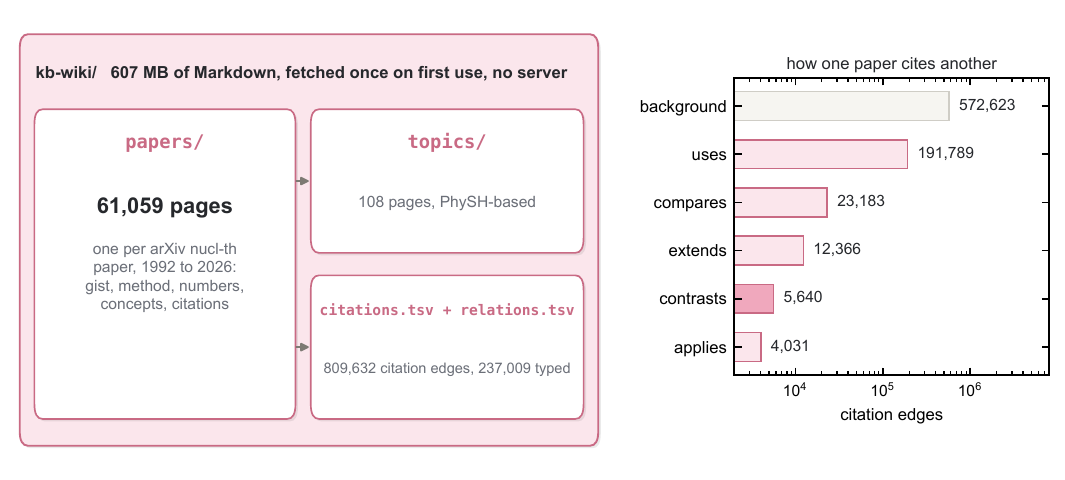}
\caption{\label{fig:kb}What ships (left) and how the 809\,632 in-corpus citations divide by type (right). A \emph{contrasts} edge records disagreement with the cited paper and survived a focused second-pass check.}
\end{figure}

FUSION also provides a searchable literature collection, downloaded once on first use: one page for each of 61\,059 arXiv nucl-th papers from 1992 to 2026, 108 topic pages based on the APS Physics Subject Headings (PhySH), and 809\,632 citation links inside the collection, of which 237\,009 are labelled by how the citing paper uses the cited one beyond plain background (Fig.~\ref{fig:kb}). The agent reads all of it with ordinary text search, guided by the \texttt{kb-search} skill. There is no server, no database and no embedding index, and after the download no network access.

\begin{figure}[!htb]\centering
\includegraphics[width=0.92\textwidth]{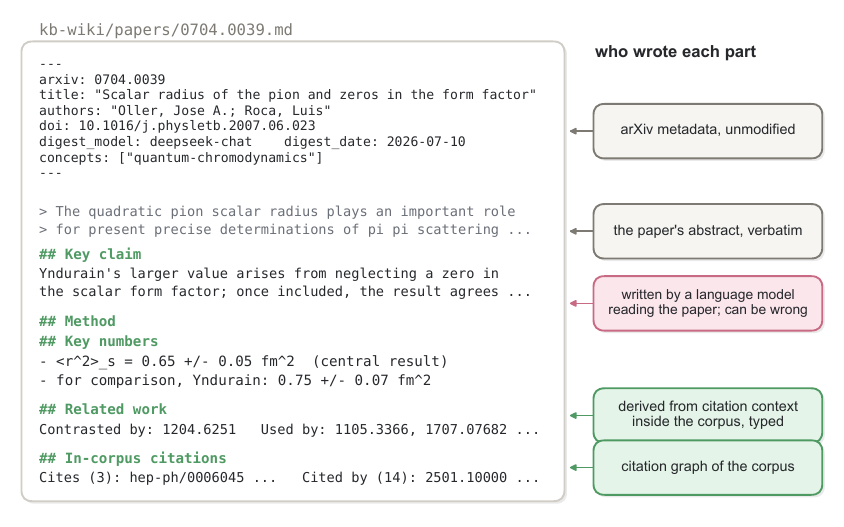}
\caption{\label{fig:page}One page of the knowledge base, with the origin of each part. Only the metadata and the abstract are the paper's own; the digest sections are model-written and say so in the header; the citation sections are computed from the corpus.}
\end{figure}

Figure~\ref{fig:page} shows what a page contains and, more to the point, who wrote each part. The metadata are arXiv's own, which it makes available for reuse, and the abstract is quoted verbatim. Everything below them was written by a language model reading the paper, and the model and date are recorded on every page, so a summary can always be traced to how it was made. No paper source, PDF, or full-text database is redistributed; every page links back by arXiv identifier and DOI, and an author who wants their page removed can ask.

Two properties matter more than the size. Plain text fails conservatively: a lexical search may miss a relevant paper, but it does not replace a miss with a confident, semantically similar one, and the skill says so. And the digests are machine-generated and some are wrong, so they are a way to find things, not a source to quote; the instruction throughout is to cite the paper, never the page. A recorded task shows the layer working alone. From a single PDF, fully offline with no external call, the corpus resolved the arXiv identifier, returned the eight corpus papers it cites and the five that cite it, and supplied digest numbers that could be checked against the paper. Citation edges come only from identifiers the citing paper itself carries (arXiv numbers and DOIs in its source) and from its INSPIRE reference list, and are counted only inside the corpus, so an edge is a lead to check against the citing paper's bibliography, and a cited-by count from it understates the true one.

\section{A real run}\label{sec:run}

\begin{figure}[!htb]\centering
\includegraphics[width=0.98\textwidth]{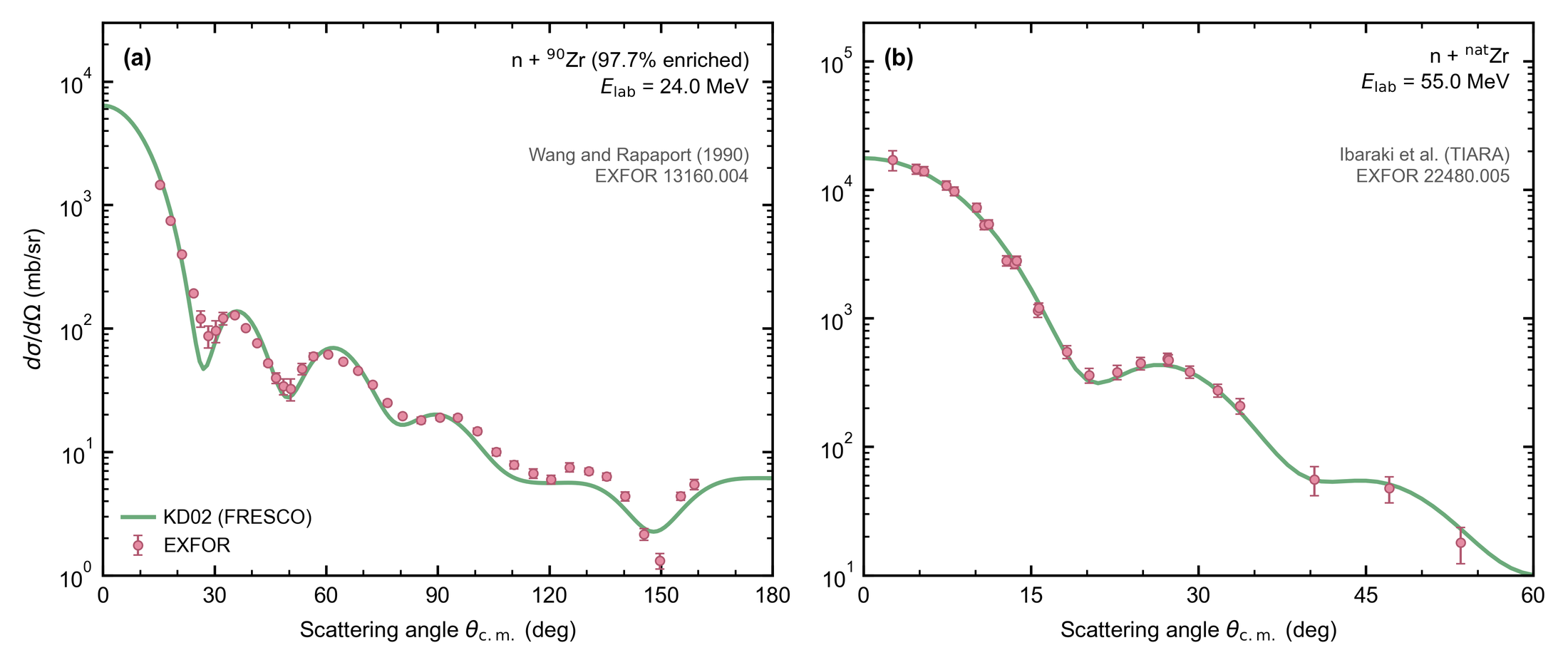}
\caption{\label{fig:case}The figure the session produced. Green is the zero-free-parameter KD02 prediction computed with FRESCO through the skill; pink points are EXFOR measurements. Left: 24~MeV, enriched $^{90}$Zr, full angular range (Wang and Rapaport, EXFOR 13160.004). Right: 55~MeV, natural Zr, forward angles (Ibaraki et al., TIARA, EXFOR 22480.005). There is no 50~MeV panel because no measurement exists at that energy.}
\end{figure}

Figure~\ref{fig:case} comes from a recorded session, with the commands and numbers as recorded. The request was one sentence: compute n+$^{90}$Zr elastic scattering at 50~MeV with the Koning-Delaroche global optical potential~\cite{Koning2003} (KD02 below), then compare against whatever EXFOR has.

The agent loaded the FRESCO skill, read the KD02 parameters from Koning's own Fortran implementation of the potential (\texttt{kd02.f}), already on disk, instead of reciting them from memory, and wrote the deck. It caught a trap that fails silently: FRESCO builds radii as $R = r_0 (A_p^{1/3} + A_t^{1/3})$ while KD02 is defined on $R = r_0 A_t^{1/3}$, so unless the projectile term is switched off (\texttt{ap=0} in the deck) every radius is 22\% too large and the resulting cross section still looks entirely plausible. The reaction cross section came out at $\sigma_R = 1301.640$~mb, moving by $1.8\times10^{-6}$ in relative terms when the radial step was halved, the partial-wave limit doubled and the truncation dropped. A second, structurally unrelated solver then checked it: COLOSS, a complex-scaled Lagrange-Laguerre code, gave 1299.1905~mb against FRESCO's 1299.19061~mb, the same number on macOS and on Linux, and a third, the Runge-Kutta solver inside the BAND scattering emulator~\cite{Odell2024}, gave 1299.1905~mb on the same input. They differ by about $10^{-4}$~mb on 1300, far below the three or four digits a global optical potential carries. Agreement at this level tests that the solvers integrate the same problem; it says nothing about whether any of them is accurate. The offset from the first number is larger and physical, 2.45~mb between integer and physical masses, and belongs to neither solver. Both decks and the convergence study are in the repository.

EXFOR contains no n+$^{90}$Zr elastic measurement at 50~MeV. The agent reported that absence as part of the result, repeated the calculation at the two energies that were measured, and produced Fig.~\ref{fig:case} with no free parameters: the mean calculated-to-measured ratio is 0.929 at 24~MeV and 0.995 at 55~MeV.

\section{Availability and status}

FUSION is released under the MIT license at \url{https://github.com/jinleiphys/FUSION}, with documentation at \url{https://vibeinscience.com}. This license covers the platform and its skills, not the physics codes. Those remain under their authors' terms, several of them GPL and one non-commercial. A skill is a directory of text and scripts with a plugin manifest for each agent; installation has been exercised in Claude Code, Codex, Antigravity, the Gemini command line and opencode, and the adversarial reviews of Sec.~V were run from one of them.

Version 0.3.0 is used daily by its author but has not yet been tested by independent users. Installation on a clean machine is the least tested path, and the one audit of it is worth reporting with its edges. All twenty install scripts were run on a Linux box that had a compiler but none of GSL, FFTW, HDF5, Boost, Eigen or Julia. Eight installed and passed the skill's own verification unattended. Four stopped with a message naming the dependency they needed, GSL, HDF5, Julia or MPI, which is what they should do. Five failed on portability defects and passed once those were fixed, one of them only after the verification budget was raised to the runtime its own skill documents. For the last three the limit was the test harness, not the skill: twice it ran no verification step at all, and once it returned an error in zero seconds on a case whose benchmark then ran by hand in under a minute. The audit found six portability defects in all, the sharpest an install probe that had always passed a zero divisor, silently harmless on Apple Silicon, where the architecture defines division by zero, and SIGFPE on x86-64. TALYS is the largest optional download at about 11~GB.

The most useful report is a calculation that passed a skill's checks but was physically wrong. Build failures, including the error and platform, are also useful, as are requests for codes not yet covered. Reports in Chinese or English may be filed through the repository. Independent use is now the missing test.

\FloatBarrier
\begin{acknowledgments}
I thank P.~Descouvemont for the suggestion that this work be described publicly. This work was supported by the National Natural Science Foundation of China (Grant Nos.~12475132 and 12535009) and the Fundamental Research Funds for the Central Universities. The skills, the knowledge base and this manuscript were developed with substantial assistance from large language models; every benchmark number quoted here was produced by running the code named beside it, and every citation was verified against the publisher record.
\end{acknowledgments}

\end{document}